\documentclass[epsf]{aastex}

\include{epsf}
\usepackage{emulateapj5}

\shorttitle{Bulge Globular Clusters}
\shortauthors{Forbes, Brodie \& Larsen}

\def\etal{{\it et al. }}

\begin{document}


\title{Bulge Globular Clusters in Spiral Galaxies}


\author{Duncan A. Forbes}
\affil{Astrophysics \& Supercomputing, Swinburne University,
  Hawthorn, VIC 3122, Australia}
\email{dforbes@swin.edu.au}

\author{Jean P. Brodie and S{\o}ren S. Larsen}
\affil{Lick Observatory, University of California,
 Santa Cruz, CA 95064, USA}
\email{brodie, soeren@ucolick.org}


\begin{abstract}
There is now strong evidence that the metal-rich globular
clusters (GC) near the center of our Galaxy are associated with the
Galactic bulge rather than the disk as previously thought. 
Here we extend the concept of bulge GCs 
to the GC systems of nearby spiral galaxies. 
In particular, the kinematic and 
metallicity properties of the GC systems favor
a bulge rather than a disk origin. 
The number of metal-rich GCs normalized by the bulge luminosity
is roughly constant (i.e. bulge S$_N$ $\sim$ 1) 
in nearby spirals, and this value is similar to that for field
ellipticals when only the red (metal--rich) GCs are considered. 
We argue that the metallicity distributions of GCs in spiral and
elliptical galaxies are remarkably similar, and that they obey the same
correlation of mean GC metallicity with host galaxy mass. 
We further suggest that the metal-rich GCs in spirals are the
direct analogs of the red GCs seen in ellipticals. The
formation of a bulge/spheroidal stellar system is  
accompanied by the formation of metal-rich GCs. 
The similarities between GC systems in spiral
and elliptical galaxies appear to be greater than the differences.

\end{abstract}


\section{Introduction}\label{sec_intro}

Globular clusters (GCs) in our Galaxy can be broadly divided into two
classes on the basis of their metallicity and/or kinematics  
(e.g. Zinn 1985). The metal-poor, non-rotating subpopulation has
long been associated with the Galaxy halo. The metal-rich GC
system 
reveals significant rotation and has historically
been associated with the disk. 
Following early suggestions by Harris (1976), 
a view is now emerging that metal-rich GCs within $\sim$ 5 kpc 
of the Milky Way 
galactic center are associated with the bulge rather than the disk 
(Frenk \& White 1982; Minniti 1995; Cote 1999). Specifically, 
the central 
metal-rich GCs are spherically distributed about the galaxy
center and overlap in metallicity with the bulge field
stars. In terms of kinematics, the GCs have a similar 
velocity dispersion and 
reveal solid body rotation matching that of the bulge stars. 
Beyond $\sim$ 5 kpc  
the metal-rich GCs have properties consistent with the thick disk 
component. 
The GCs also appear to be coeval with the bulge stars (Ortolani
\etal 1995). 
Cote \etal (2000) has sucessfully modelled the Galactic GC system
based on hierarchical build-up around a protobulge and its 
metal-rich GCs.

Here we extend this view of `bulge GCs' to other spiral galaxies. In
particular, we suggest that the inner metal-rich GCs in M31, M81
and M104 spiral galaxies  
are associated with their bulges. We
further suggest that the red (metal-rich) GC subpopulations in 
giant ellipticals are their analogs, in agreement with the view
advocated by Cote \etal (2000). Thus the bulges of
spirals, S0s
and the entire stellar component of ellipticals (which we 
collectively refer to as the `bulge') 
may all have associated metal-rich GCs. We briefly discuss the
implications for GC and galaxy formation.

\section{Bulge Globular Cluster Kinematics}

\subsection{M31}

The Andromeda galaxy (M31; Sb) reveals a 
bimodal GC metallicity distribution with the metal-rich GCs
preferrentially close to the galaxy center  
(e.g. Huchra, Kent \& Brodie 1991).  
Based on their kinematics, Huchra \etal   
concluded that 
interior to $\sim$2 kpc the 
metal-rich GCs were rapidly rotating.
Recently Perrett \etal (2001) have obtained
velocities for over 200 GCs in the M31 system. They find a 
velocity dispersion of  146 $\pm$ 12 km/s. This is
consistent with the central {\it stellar} velocity dispersion of 150
km/s (van den Bergh 1999). Furthermore the metal-rich GCs reveal
solid-body-like rotation within 5 kpc with 
an amplitude similar to that of the
stellar rotation curve which is dominated by the bulge at these
small radii (e.g. Rubin \& Ford 1970). They also find
that the metal-rich
GC system is spherically distributed about the galaxy center.
Thus, 
the inner metal-rich GCs in M31 reveal the same features that
have lead previous workers to associate equivalent GCs in our
Galaxy with the bulge. 

\subsection{M81}

The GC system of M81 (Sa/Sb) is less well studied than M31
but both photometric (Perelmuter \& Racine 1995) and kinematic
studies (Perelmuter, Brodie \& Huchra 1995; 
Schroder \etal 2001) have noted similarities to the
Milky Way's GC system. 
Schroder
\etal (2001) give the kinematics for the metal-rich GCs within 2
kpc of the galaxy center from Keck spectra.  
They derive a velocity dispersion of 152 
$\pm$ 36 km/s and rotation velocity of 96 $\pm$ 56
km/s. Measurements of the {\it stellar} central velocity
dispersion vary from 150 to 180 km/s with a median value of 167
km/s. The {\it stellar} rotation curve for M81 peaks at around 0.5 kpc
radius with a value of $\sim$110 km/s (Heraudeau \& Simien 1998).   
The stellar values are slightly higher than those inferred for the
GC system, but are well within the errors. Although the evidence
is less strong, the inner metal-rich GCs of M81 have kinematic
properties that are consistent
with the bulge.\\ 

\section{Globular Cluster Metallicities}

Individual metallicities are now available for over 250 GCs in
M31 (Barmby \etal 2000). 
For M81, the photometry of Perelmuter \& Racine (1995) was
insufficient to clearly differentiate the metal-poor and metal-rich
subpopulations but they did note that the inner GC sample was
dominated by red (metal-rich) objects. 
Perhaps the best photometrically-studied GC system in a spiral
galaxy beyond the Local Group is that of the Sombrero galaxy (M104; Sa).
The recent HST imaging study of Larsen,
Forbes \& Brodie (2001) showed that the GC system has two
distinct subpopulations. After Galactic extinction correction and 
color transformation using 
Kissler--Patig \etal (1998), we show in Fig. 1 
the metallicity distribution of GCs in
M104 compared to that of M31,
the Milky Way and M33 (data from Forbes \etal 2000). 
If we consider a cut at say [Fe/H] = --1, then it is clear that the 
ratio of metal-rich to metal-poor GCs is significantly higher in M104 
than the other spirals. 
Since M104 has a tiny disk and a dominant 
bulge it is tempting to associate 
the metal-rich GCs with the bulge rather than the disk component 
(see also Section 4).  
If the samples are restricted to the GCs to within 5 kpc of the galaxy
center then the situation is even more pronounced.

M33 (Sc) at the opposite extreme has very 
few GCs with [Fe/H] $>$ --1. Its `bulge' 
has a luminosity of M$_V$ $\sim$ --15 (Bothun 1992).
Local Group dwarf galaxies
of this luminosity, typically have less than half a dozen GCs. 
Thus if the M33 bulge is analogous to a
small galaxy few associated GCs are expected. We note that
the LMC (another bulgeless galaxy) also lacks metal-rich GCs. 
For M31 (Sb) and the Milky Way (Sbc) the number of metal-rich
GCs (relative to metal-poor ones) 
is intermediate between M104 (Sa) and M33 (Sc). 
We note that Kissler-Patig \etal (1997) associate the 
red GCs in the S0 galaxy NGC
1380 with that galaxy's bulge. {\it It
seems likely that the relative number of metal-rich
GCs is related to the host galaxy Hubble type and hence the
relative importance of a galaxy's bulge. }

How are the metal-rich and metal-poor
subpopulations in spiral galaxies related to those seen in early
type galaxies ? 
Historically, one difference between 
the GC metallicity distributions 
in spirals and ellipticals were thought to be the mean 
metallicity of the two peaks. 
M104, M31 and the Milky Way all have GC subpopulations with 
mean metallicities of [Fe/H] $\sim$ --1.5 and --0.5 (see Fig. 1) 
while ellipticals were thought to have GC mean 
values of [Fe/H] $\sim$ --1.0
and 0.0 (Harris 1991). 
Recently, two developments 
have caused us to reassess the mean GC metallicity 
in ellipticals towards lower metallicities. 
The first effect is the use of 
more accurate transformations from optical colors
to [Fe/H]. For example, the new 
transformation of 
Kissler--Patig \etal (1998) converts a typical V--I = 1.05 
to [Fe/H] = --1.07, where the old Galactic--based transformation would 
give [Fe/H] $\sim$ --0.5 (Couture \etal 1990).
The second effect is that the more accurate 
Galactic extinction values of Schlegel \etal (1998) tend to be larger on 
average by up to A$_V$ $\sim$ 0.1, than the 
traditionally-used Burstein \& 
Heiles (1984) values. Thus extinction-corrected GC colors are now bluer 
than before, and more metal-poor when transformed. 
If these two effects are taken into account, the two GC subpopulations 
in ellipticals have 
mean metallicities of [Fe/H] $\sim$ --1.5 and --0.5 
which is similar to those in spirals. 
{\it To first order there appears
to be very little difference between the mean metallicity of the
two subpopulations in late and early-type galaxies.  }

\vbox{
\begin{center}
\leavevmode
\hbox{%
\epsfxsize=9cm
\epsffile{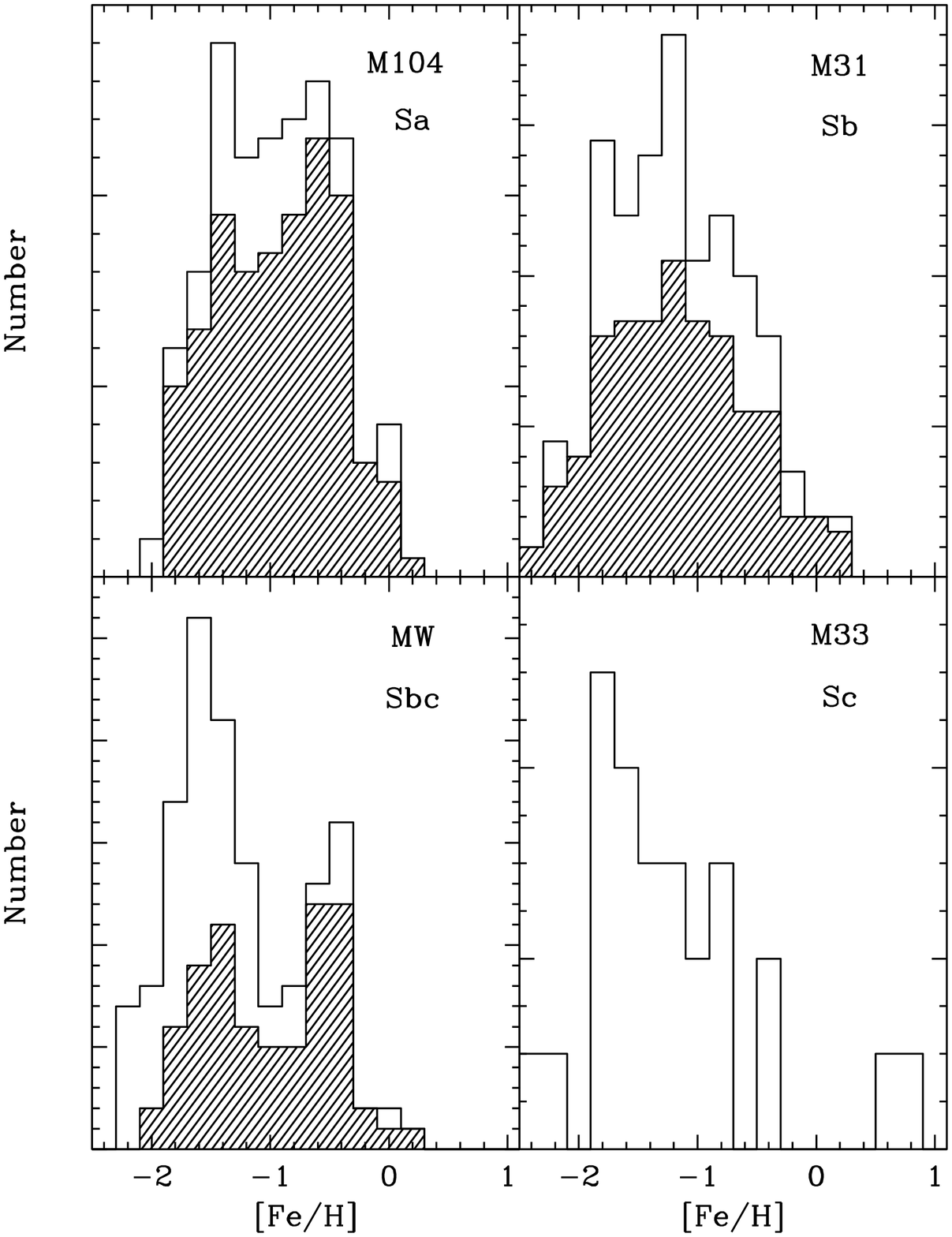}}
\figcaption{\small
Metallicity distributions of GCs in M104, M31, Milky Way and
M33. The y axis has been scaled arbitrarily. 
Open histograms show the total GC population observed, and
hashed histograms show GCs with 5 kpc of the galactic center
(except for M33).  
M104 has more metal-rich (e.g. [Fe/H] $>$ --1) to metal-poor  
GCs than the later type spirals.  
\label{fe}}
\end{center}}

When GC metallicities are examined in still more detail, it is
found that the mean GC color (metallicity)
correlates with galaxy velocity dispersion  for {\it early-type}
galaxies (Forbes \& Forte 2001; Larsen \etal 2001). 
Do the bulge GCs of spirals follow the same 
relation as early-type galaxies ?  

We have collected a sample of 37 early-type 
galaxies from the literature with bimodal GC
color distributions. The mean color of the metal-rich
subpopulation has been corrected to
a common V--I color (Forbes \& Forte 2001) and corrected for extinction
using Schlegel \etal (1998). 
To this sample we add M104, M31 and the 
Milky Way GC systems (the combined sample data are available at 
http://astronomy.swin.edu.au/staff/dforbes/glob.html). The V--I color of
the M31 and Milky Way metal-rich GCs have been
calculated using the transformations of Barmby \etal
(2000). Central velocity dispersions come from Gebhardt \etal
(2000) and Kent (1992). 
The uncertainty in the mean color 
is rarely quoted in the original works. 
We have decided to adopt 
relatively conservative error estimates (i.e $\pm$ 0.03$^m$ for HST data 
with definite bimodality, $\pm$ 0.05 for probable bimodality 
and $\pm$ 0.08 
for ground-based data to reflect the higher photometric errors and 
contamination rates). 
They
may be smaller than we assume since 
the scatter in the data points is generally less than the
errors. 
This means that we will tend to
underestimate the significance of any slope compared to the error
on the slope from a least-squares fit. 

\vbox{
\begin{center}
\leavevmode
\hbox{%
\epsfxsize=9cm
\epsffile{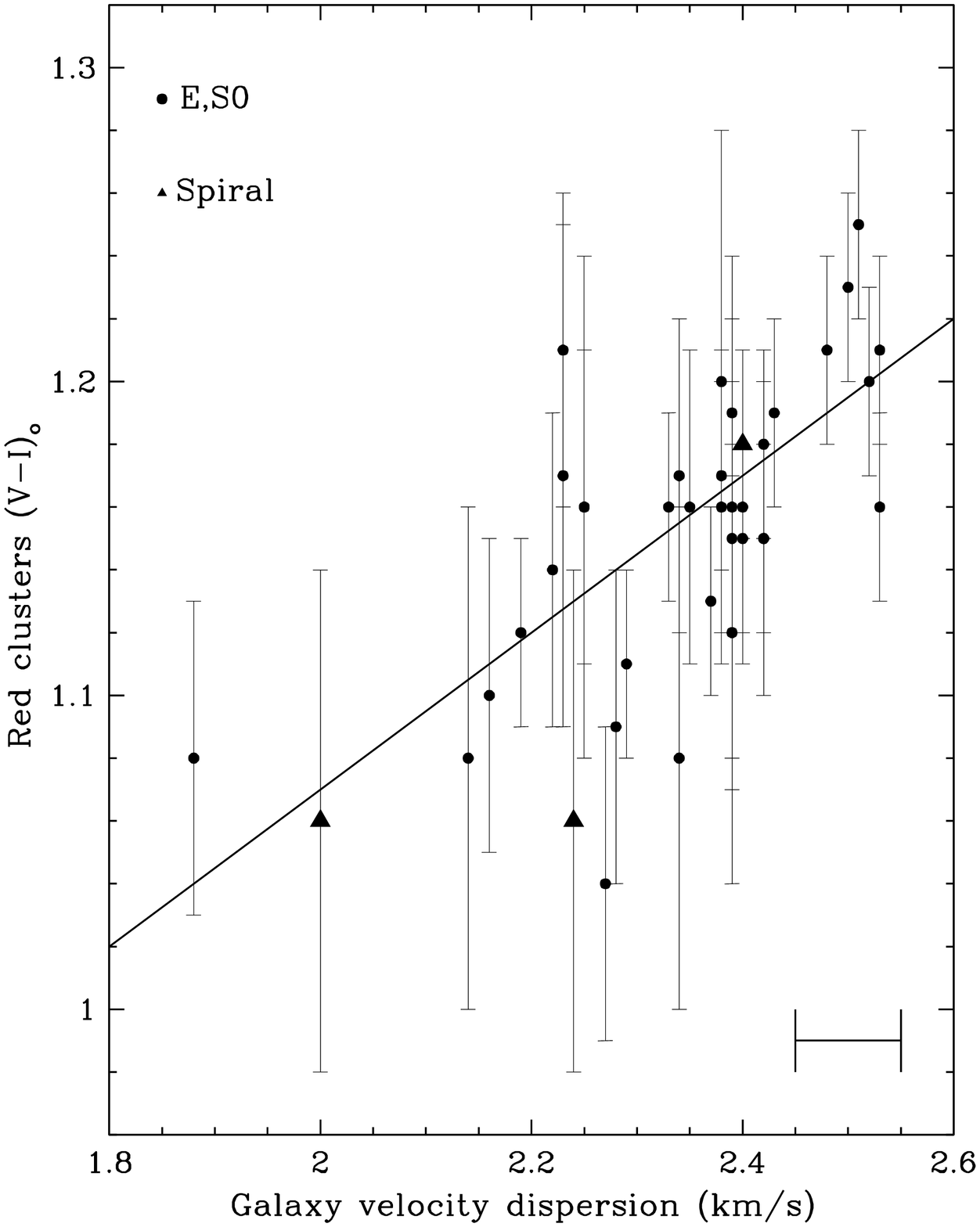}}
\figcaption{\small
Mean color of the red (metal-rich) 
globular cluster subpopulations 
versus log galaxy velocity dispersion. Early-type galaxies are shown
by filled circles and spirals by triangles. 
A typical velocity dispersion error is shown in the lower left. 
The solid line shows the best fit to the early-type galaxies
(slope = 0.26$\pm$0.06, intercept = 0.56$\pm$0.14). The
correlation is significant at the 
4$\sigma$ level.
The three spiral galaxies (MW, M31, M104) 
are consistent with the overall early-type galaxy
relation.
\label{red}}
\end{center}}

The data are shown in Fig. 2. For the early-type galaxies,  
the Spearman rank correlation indicates that 
the red GCs are correlated with galaxy velocity dispersion with
a probability of 99.9\%.
A least squares fit gives a positive slope (similar to that found
by Forbes \& Forte (2001) and Larsen \etal 
(2001) for smaller samples) at the
4$\sigma$ level. 
The mean colors of the metal-rich 
GC subpopulations in the three spirals are
also plotted in Fig. 2. {\it The red GCs in spirals 
are consistent with the metallicity -- velocity dispersion 
relation for early-type
galaxies.}

If we use only the high quality sample (i.e. HST data with 
definite bimodality) then the Spearman test gives 99.5\% and 
a slope of 4$\sigma$.
An unweighted fit to the high quality sample gives a similar
slope, with slightly increased significance of 5$\sigma$.

\section{Bulge Specific Frequency}

Traditionally GC specific frequency S$_N$ refers to the total number of
GCs per galaxy luminosity, normalised to M$_V$ = --15. Harris
(1981) was the first to compare {\it total} GC numbers in spirals
with the the luminosity of the bulge component. Recently, Cote
\etal (2000) pointed out that S$_N$ defined in this manner was 
indistinguishable for spirals and ellipticals in similar
environments. 
Here we focus on
the metal-rich/red Gcs in spirals/ellipticals, compared to the
bulge luminosity (we assume that ellipticals are bulge dominated
systems). We refer to this as the bulge S$_N$. 

The total number of 
GCs and the number of metal-rich GCs are given by 
Larsen, Forbes \& Brodie (2000) for M104, and in the compilation
of Forbes \etal
(2000) for M31 and the Milky Way. These numbers and the
host galaxy magnitudes discussed below are summarized in Table 1. 
From the galaxy total magnitudes, we calculate the bulge and disk
magnitudes using the following method.  
The bulge-to-total (B/T) luminosity for M104 has been given by Kent
(1988) as 0.85 and by Baggett \etal (1998) as 0.73. Here we use 
0.8. For M31 (Sb) and the MW (Sbc) we use the
B/T variations with Hubble type of Simien \& de
Vaucouleurs (1986), i.e. B/T = 0.25 and B/T = 0.19
respectively with a dispersion of about $\pm$0.05 within a given Hubble 
type. 
For the disk contribution we assume that the halo light is negligible
and hence all of the remaining light comes from the disk. 

In Section 2 we argued, mostly from the kinematic data, that
the bulk of metal-rich GCs in spirals are not associated with the disk
but rather the bulge component. Further support for this
idea comes from examining the number of metal-rich GCs per unit
starlight. The GC system of M104 provides a key
data point. 
For M104 the total number of red GCs and the
disk magnitude 
combine to give a disk S$_N$ of
4.4$\pm$5.2 Assuming that the disks in M31 and the Milky Way 
have similar stellar populations (i.e. M/L), 
the disk S$_N$ of M104 is about 20
times that of these other spirals. 
This large variation in disk S$_N$ suggests that 
that the bulk of metal-rich GCs in spirals are not in fact disk objects. 

From the bulge magnitudes and number of metal-rich GCs given
above, we derive bulge S$_N$ values of 1.1$\pm$0.8 (M104), 
0.6$\pm$0.3 (M31) and 0.8$\pm$0.9 (MW). 
Unlike the disk S$_N$ values, bulge S$_N$ values are
fairly consistent between the three spirals. 

In the case of the Milky Way only GCs within $\sim$ 5 kpc show
bulge characteristics while those 
further out have been associated with the thick disk 
(Minniti 1995; Cote 1999). In terms of
the bulge effective radius, 5 kpc is 2~R$_{eff}$
(van den Bergh 1999). 
The bulge effective radii for M104 and M31
are 8 kpc (Bender, Burstein \& Faber 1992) and 2.5 kpc (van den
Bergh 1999) respectively. 
If the metal-rich GC samples in M104
and M31 are restricted 
to within 2~R$_{eff}$, then we estimate about 378 metal-rich GCs
in M104 (from Larsen, Forbes \& Brodie 2001) and 61 in M31 (from
Barmby \etal 2000). The Milky Way has about 35 known metal-rich
GCs within 2~R$_{eff}$. 
Thus the bulge S$_N$ values within 2~R$_{eff}$ are 
0.6$\pm$0.5, 0.4$\pm$0.2 and 0.6$\pm$0.6
for M104, M31 and the Milky Way respectively. 
Within the errors, the bulge S$_N$ values for M104, M31 and the
Milky Way are consistent. {\it So although the three spirals span a range
of Hubble types from Sa to Sbc, the bulge S$_N$ appears to be  
nearly constant for spiral galaxies.} 

In each case, there is a tendency to miss some 
metal-rich GCs as they are harder to detect near galaxy
centers.
For example, about a dozen GCs are thought to
be hidden from our view in the Milky Way (van den Bergh 1999). 
Recently the 2MASS survey
has detected two more metal-rich bulge GCs (Hurt \etal 2000). 
Barmby \etal (2000) give photometry for about 2/3 of the total
GC population in M31 which has 
an estimated total population of 400 $\pm$ 55. 
Again many of the missing GCs will be associated with
the bulge. The derived bulge S$_N$ values may be 
underestimated by up to 30\%.

How do the bulge S$_N$ values for spirals compare to ellipticals
? Field ellipticals have {\it total} S$_N$ values of 1--3 (Harris
1991). The fraction of red GCs in ellipticals is typically about
half (e.g. Forbes, Brodie \& Grillmair 1997). For example, wide
area studies of the 
field/group ellipticals NGC 1052 and NGC 1700 found red fractions and
total S$_N$ values of 0.50, 1.7 and 0.56, 1.3 respectively
(Forbes, Georgakakis \& Brodie 2001; Brown \etal 2000). This
implies that field ellipticals typically have {\it bulge} S$_N$
values of 0.5--1.5. {\it Thus field ellipticals have similar
bulge S$_N$ values to field spirals.} 
This provides
further support for our claim 
that the metal-rich GCs in spirals and those in ellipticals have
the same origin, i.e. they formed along with the bulge stars. 
We note that cluster ellipticals may
have similar S$_N$ values when the mass of hot gas is taken into
account (McLaughlin 1999). Little is known about the
GC systems of cluster spirals.

\section{Concluding Remarks}\label{sec_conc}

From globular cluster (GC) 
kinematic information we have argued that the inner,
metal-rich GCs in the nearby spirals M31 and M81 have a bulge
origin. 
On the basis of GC numbers and specific frequency, 
we showed that the metal-rich GCs in the Sa spiral M104 are most
likely associated with the dominant bulge rather than
the small disk component. 
The derived bulge specific frequency for the GCs in
M104, M31 and the Milky Way are consistent with a constant value
of $\sim$1. This is similar to the value for field ellipticals (when
only the metal-rich GCs are considered) but is less than that for
cluster ellipticals. 
The metallicity distributions of GCs in late and
early-type galaxies are similar to first order, and obey the same
correlation of mean GC metallicity with host galaxy velocity
dispsersion.  
This relation, 
indicates a common chemical enrichment history for the metal-rich
GCs and the host galaxy (Forbes \& Forte 2001).

We conclude that the majority of the metal-rich GCs in spirals are
associated with the galaxy bulge, and that these GCs are the
analogs of the red (metal-rich) GCs in giant ellipticals. 
Thus
GC systems provide another example of the similarity between
ellipticals and spiral bulges (e.g. Wyse, Gilmore \& Franx 1997).
By extension this would suggest that bulges and ellipticals
formed by a similar mechanism.
In the case of the Milky Way bulge, van den
Bergh (1996) concluded that it was formed
by a rapid but clumpy collapse.  

In the multi--phase collapse
model for GC formation proposed by Forbes, Brodie \& Grillmair
(1997) the `bulge' of a giant elliptical galaxy occurred in the second
or galactic phase. The red GCs formed during this phase. In
that paper, we associated the metal-rich GCs of spirals with
disks and speculated that they were a third phase of GC
formation. It now seems likely that
the bulk of metal-rich GCs in spirals were formed along with the 
bulge stars and it is these that are 
directly analogous to red GCs in giant ellipticals.

\section{Acknowledgments}\label{sec_ack}

We thank M. Beasley, J. Huchra and L. Schroder for their comments.
We also thank the referee, P. Cote, for suggesting several
improvements to the paper. 
Part of this research was funded by the Ian Potter Foundation 
and NSF grant AST 9900732.

\begin{table}
\begin{scriptsize}
\begin{center}
\renewcommand{\arraystretch}{1.5}
\begin{tabular}{lccc}
\multicolumn{4}{c}{\scriptsize TABLE 1}\\
\multicolumn{4}{c}{\scriptsize GLOBULAR CLUSTER BULGE SPECIFIC FREQUENCY}\\
\hline
\hline
 & M104 & M31 & Milky Way\\
\hline
Globular cluster total & 1150$\pm$575 & 400$\pm$55 & 160$\pm$20\\
Metal-rich clusters (all radii) & 667$\pm$333 & 100$\pm$14 & 53$\pm$7\\
Metal-rich clusters ($<$2R$_{eff}$) & 378$\pm$189 & 61$\pm$8 & 35$\pm$4\\
Galaxy total M$_V$ & --22.2$\pm$0.1 & --22.0$\pm$0.2 & --21.3$\pm$0.3\\
Bulge-to-total ratio & 0.80$\pm$0.05 & 0.25$\pm$0.05 & 0.19$\pm$0.05\\
Disk S$_N$ (all radii) & 4.4$\pm$5.2 & 0.2$\pm$0.1 & 0.2$\pm$0.1\\ 
Bulge S$_N$ (all radii) & 1.1$\pm$0.8 & 0.6$\pm$0.3 & 0.8$\pm$0.9\\
Bulge S$_N$ ($<$2R$_{eff}$) & 0.6$\pm$0.5 & 0.4$\pm$0.2 &
0.6$\pm$0.6\\

\hline
\end{tabular}
\end{center}
\tablenotetext{a}{\scriptsize The table lists the total number of globular
clusters, the metal-rich subpopulation and those within
twice the bulge R$_{eff}$. 
The bulge and disk specific
frequencies, S$_N$, use bulge and disk luminosities
respectively. 
}
\end{scriptsize}
\end{table}


\begin{thebibliography}{} 





{\bibitem{5} Baggett, W. E., Baggett, S. M., Anderson, K. S. J.,
1998, AJ, 116, 1626}

{\bibitem{8} Barmby, P., Huchra, J., Brodie, J., Forbes, D.,
Schroder, L., Grillmair, C., 2000, AJ, 119, 727}


{\bibitem{6} Bothun, G., 1992, AJ, 103, 104}

{\bibitem{10} Bender, R., Burtsein, D., Faber, S., 1992, ApJ,
399, 462}




{\bibitem{11} Brown, R. J. N., Forbes D. A., Kissler-Patig M., Brodie J.,
2000, MNRAS, 317, 406}

{\bibitem{12} Burstein, D., Heiles, C., 1984, ApJS, 54, 33}



{\bibitem{14} Cote, P., 1999, AJ, 118, 406}

{\bibitem{15} Cote, P., Marzke, R. O., West, M. J., Minniti, D.,
2000, ApJ, 533, 869}

{\bibitem{16} Couture, J., Harris, W. E., Allwright, J. W. B.,
1990, ApJS, 73, 671} 


{\bibitem{20} Forbes D., Masters, K., Minniti, D., Barmby, P.,
2000, A\&A, 358, 471}

{\bibitem{21} Forbes, D., Georgakakis, A., Brodie, J., 2001,
MNRAS, in press}

{\bibitem{22} Forbes, D., Brodie, J., Grillmair, C.,
1997, AJ, 113, 1652} 


{\bibitem{81} Forbes, D., Forte, J., 2001, MNRAS, 322, 257}


{\bibitem{77} Frenk, C., White, S. D. M., 1982, MNRAS, 198, 173}

{\bibitem{78} Gebhardt, K., \etal 2000, ApJ, 539, L13}



{\bibitem{26} Harris, W. E., 1976, AJ, 81, 1095}

{\bibitem{25} Harris, W. E., 1981, ApJ, 251, 497}

{\bibitem{28} Harris, W. E., 1991, ARA\&A, 29, 543}

{\bibitem{29} Heraudeau, Ph., \& Simien,  F., 1998, A\&AS, 133, 317}

{\bibitem{40} Huchra, J. P., Kent, S., Brodie, J. P., 1991, ApJ,
370, 495}

{\bibitem{42} Hurt, R., \etal 2000, astro-ph/0006262}

{\bibitem{36} Kent, S., 1988, AJ, 96, 514}

{\bibitem{42} Kent, S., 1992, ApJ, 387, 181}

{\bibitem{35} Kissler-Patig, M., Richtler, T., Storm, J., 
della Valle, M., 1997, A\&A, 327, 503}

{\bibitem{40} Kissler-Patig, M., Brodie, J., Schroder, L.,
Forbes, D., Grillmair, C., Huchra, J., 1998, AJ, 115, 105}



{\bibitem{38} Larsen, S., Forbes, D., Brodie, J., 2001, MNRAS,
submitted}

{\bibitem{39} Larsen, S., Brodie, J., Huchra, J., Forbes, D.,
Grillmair, C., 2001, AJ, in press}

{\bibitem{37} McLaughlin, D., 1999, AJ, 117, 2398}

{\bibitem{41} Minniti, D., 1995, AJ, 109, 1663}


{\bibitem{37} Ortolani, S., \etal 1995, Nature, 377, 701}

{\bibitem{97} Perelmuter, J-M., Brodie, J. P., Huchra, J. P.,
1995, AJ, 110, 620}

{\bibitem{98} Perelmuter, J-M., Racine, R., 1995, AJ, 109, 1055}

{\bibitem{99} Perrett, K., \etal 2001, in preparation}



{\bibitem{49} Rubin, V., Ford, W., 1970, ApJ, 159, 379}

{\bibitem{88} Schlegel, D. J., Finkbeiner, D. P., Davis, M.,
1998, ApJ, 500, 525}


{\bibitem{49} Simien, F., de Vaucouleurs, G., 1986, ApJ, 302, 564}


{\bibitem{51} Schroder, L., Brodie, J., Kissler-Patig, M.,
Huchra, J., Phillips, A., 2001, AJ, in press}




{\bibitem{54} van den Bergh, S., 1999, A\&AR, 9, 273}



{\bibitem{63} Wyse, R., Gilmore, G., Franx, M., 1997, ARAA, 35, 637}

{\bibitem{64} Zinn, R., 1985, ApJ, 293, 424}

\end{thebibliography}
\end{document}